\documentclass[prd,aps,nofootinbib,floatfix,10pt]{revtex4}
\usepackage{amsmath,graphicx,epsfig,amssymb,dsfont,mathtools,}
\usepackage[usenames]{color}
\usepackage{ulem} 
\usepackage{cancel}
\usepackage{bigstrut}
\usepackage{slashed}
\usepackage{multirow}
\usepackage{makecell}
\usepackage{verbatim}
\usepackage{amsmath,graphicx,color,epsfig}
\usepackage{longtable}
\usepackage{subfigure}
  \allowdisplaybreaks[1]

\allowdisplaybreaks

\begin{document}
\title{Bottom baryon decays into light meson and dark baryon within perturbative QCD approach}
\author{Ye Xing}
\email{Corresponding author. xingye_guang@cumt.edu.cn}
\affiliation{School of Material Science and Physics, China University of Mining and Technology, Xuzhou 221000, China}

\author{Yu-Ji Shi}
\email{Corresponding author. shiyuji@ecust.edu.cn}
\affiliation{School of Physics, East China University of Science and Technology, Shanghai 200237, China}

\author{Xiao-hui Hu}
\email{Corresponding author. huxiaohui@cumt.edu.cn}
\affiliation{School of Material Science and Physics, China University of Mining and Technology, Xuzhou 221000, China}
\affiliation{Lanzhou Center for Theoretical Physics and Key Laboratory of Theoretical Physics of Gansu Province,
Lanzhou University, Lanzhou 730000, China}

\begin{abstract}
In the work, we study the production of dark baryon $\psi$ with the bottom-baryon decays $\Lambda_b\to M \psi$ under the perturbative QCD approach. For the Type-I/II models in B-Mesogenesis scenarios, we calculate the form factors for bottom-baryon decays into meson. Using these results, we obtain branching ratios for $\Lambda_b\to M \psi$ decays. The PQCD approach calculations show these branching ratios reach $\mathcal{O}(10^{-5})$ in both models, with $\mathcal{B}(\Lambda_b\to K\psi)$ specifically reaching $\mathcal{O}(10^{-4})$ in Type-II. These accessible magnitudes make the predictions testable at the LHCb and B-factories with improved precision.
\end{abstract}

\maketitle

\section{Introduction}
\label{sec:introduction}

The Standard Model (SM) of particle physics and the standard cosmological model ($\Lambda$CDM) constitute two remarkably successful theoretical frameworks, describing the fundamental interactions of elementary particles and the large-scale evolution of the Universe, respectively. However, their mutual incompatibility raises profound questions, including the origin of dark matter and the observed matter-antimatter asymmetry. Since Sakharov first established the necessary conditions for baryogenesis \cite{Sakharov:1967dj}, extensive efforts have been devoted to identifying viable mechanisms capable of explaining these phenomena. Conventional approaches typically rely on high-energy scales and the existence of beyond-SM particles with extreme masses, rendering them experimentally inaccessible in the foreseeable future.

A recently proposed $B$-Mesogenesis scenario \cite{Elor:2018twp,Alonso-Alvarez:2021qfd,Elahi:2021jia} presents a compelling alternative, offering a unified resolution to both the dark matter relic abundance and baryon asymmetry. In the $B$-Mesogenesis scenario, decays of a heavy scalar field $\Phi$ in the early Universe generate $b$-$\bar{b}$ quark pairs that subsequently hadronize into $B$-mesons. The neutral $B^0$/$\bar{B}^0$ mesons undergo CP-violating oscillations before decaying asymmetrically into: (i) dark sector baryons ($\mathcal{B}=-1$) and (ii) visible hadrons ($\mathcal{B}=+1$), thereby producing the observed baryon asymmetry while preserving total baryon number conservation. This framework naturally connects baryogenesis to DM production through experimentally accessible $B$-meson phenomenology.
A distinctive feature of this framework is its testability: predictions can be directly probed at hadron colliders and $B$-factories \cite{Alonso-Alvarez:2021qfd,Borsato:2021aum}, as well as indirectly constrained through precision measurements at Kaon and Hyperon facilities \cite{Alonso-Alvarez:2021oaj,Goudzovski:2022vbt}. Currently, dedicated experimental searches for $B$-Mesogenesis signatures---particularly $B$ meson decays into baryons accompanied by missing energy---are being conducted by the Belle-II \cite{Belle:2021gmc} and LHCb \cite{Rodriguez:2021urv} collaborations.

Recent theoretical studies have extensively investigated $B$-meson decays within the $B$-Mesogenesis framework, encompassing both exclusive and semi-inclusive processes. The exclusive decay $B \to p \psi$ was initially analyzed via leading-twist light-cone sum rules (LCSR) \cite{Khodjamirian:2022vta}, with higher-twist corrections subsequently computed \cite{Boushmelev:2023huu}. A comprehensive treatment of $B$-meson decays to octet baryons or charmed anti-triplet baryons plus $\psi$ was presented in \cite{Elor:2022jxy}, while analogous exclusive decays to baryons with missing energy were explored in \cite{Dib:2022ppx} for neutralino detection. Additionally, $\Lambda_b$ decays to light mesons and dark baryons have been studied using both QCD factorization \cite{Li:2024htn} and LCSR methods \cite{Shi:2024uqs}. Beyond exclusive channels, semi-inclusive decays $B \to X_{u/c,d/s} \psi$ were investigated via heavy quark expansion (HQE) in \cite{Shi:2023riy}, yielding constraints on the $B$-Mesogenesis coupling parameters.
Previous LCSR studies of $B$-meson decays in $B$-Mesogenesis have uncertainties from threshold and Borel parameters. In this work, we will use perturbative QCD (pQCD) with $k_T$-factorization to improve precision in predicting the exclusive decay channel $B \to p \psi$. This approach which has been successfully applied to the meson and baryon system~\cite{Xing:2019xti,Han:2022srw,He:2006vz,He:2006ud,Li:2010bb,Li:2008tk,Rui:2022sdc}, systematically accounts for transverse momentum effects and provides a more reliable theoretical framework for evaluating the decay amplitudes.

This paper is organized as follow: In Sec.II, we introduce the wave functions of $\Lambda_b$, $\pi$, $K$, and $D$ in turn, while Sec.III discuss the B-Mesogenesis scenario. Sec.IV contains our perturbative calculation within the PQCD framework. In Sec.V, we study the numerical results, and a conclusion is presented in the last section.

\section{Wave functions}
\label{sec:wave functions}

In general, the Lorentz structure of $\Lambda_b$ baryon wave function can be simplified using the Bargmann-Wigner equation~\cite{Hussain:1990uu} in the heavy quark limit, where the spin and orbital degrees of freedom of the light quark system are decoupled. In coordinate space, the wave function of the $\Lambda_b$ baryon is defined as~\cite{Lu:2009cm,Loinaz:1995wz},
\begin{eqnarray}
\langle0|u_{\alpha}^{iT}(x)d_{\beta}^j(y)b_{\gamma}^{k}(0)|\Lambda_b(p_{\Lambda_b})\rangle&=&
\frac{1}{6}\varepsilon^{ijk}\Big\{f_{\Lambda_b}^{(1)}\big[(\frac{\bar{\mathbf{n}\!\!\!/}\mathbf{n}\!\!\!/}{8}\phi_3^{+-}(x_2,x_3)+\frac{\mathbf{n}\!\!\!/\bar{\mathbf{n}\!\!\!/}}{8}\phi_3^{-+}(x_2,x_3))\gamma_5 C^T \big]+f_{\Lambda_b}^{(2)}\big[(\frac{\mathbf{n}\!\!\!/}{4}\phi_2(x_2,x_3)\notag\\
&&+\frac{\bar{\mathbf{n}\!\!\!/}}{4}\phi_4(x_2,x_3))\gamma_5 C^T \big]\Big\}_{\alpha\beta}  \Big\{u_{\Lambda_b}(k_i,\mu)\Big\}_{\gamma}.
\end{eqnarray}
where the light-cone vectors $\textbf{n}=1/\sqrt2(1,0,0,1)$ and $\mathbf{\bar n}=1/\sqrt2(1,0,0,-1)$, satisfying $\mathbf{n}\cdot \bar{\mathbf{n}}=1$. Here $b$, $u$, and $d$ denote quark fields, $i$, $j$, and $k$ are color indices, $\alpha$, $\beta$, and $\gamma$ are spinor indices, $C$ is the charge conjugation matrix,  and $u_{\Lambda_b}$ represents the $\Lambda_b$ baryon spinor. The normalization constant corresponds to the wave function at the origin in the configuration space. The numerical value $f_{\Lambda_b}^{(1)}\approx f_{\Lambda_b}^{(2)}\equiv f_{\Lambda_b}=0.021\pm 0.004$ $\text{GeV}^3$  is adopted from experimental data on the semileptonic decay $\Lambda_b\to \Lambda_c l\bar \nu_{l}$. The light-cone distribution amplitudes (LCDAs) $\phi_2$ ,$\phi_3^{+-}$, $\phi_3^{-+}$ and $\phi_4$ are expanded in Gegenbauer polynomials. This expansion is derived from QCD sum rules including only the leading-order perturbative contribution,
\begin{eqnarray}
&&\phi_2(x_2,x_3)=m_{\Lambda_b}^4 x_2x_3\Big[\frac{1}{\epsilon_0^4}e^{-m_{\Lambda_b}(x_2+x_3)/\epsilon_0}+a_2 C_2^{3/2}\big(\frac{x_2-x_3}{x_2+x_3}\big)\frac{1}{\epsilon_1^4}e^{-m_{\Lambda_b}(x_2+x_3)/\epsilon_1}\Big],\notag\\
&&\phi_3^{+-}(x_2,x_3)=\frac{2m_{\Lambda_b}^3x_2}{\epsilon_3^3}e^{-m_{\Lambda_b}(x_2+x_3)/\epsilon_3},\notag\\
&&\phi_3^{-+}(x_2,x_3)=\frac{2m_{\Lambda_b}^3x_3}{\epsilon_3^3}e^{-m_{\Lambda_b}(x_2+x_3)/\epsilon_3},\notag\\
&&\phi_4(x_2,x_3)=\frac{5}{\mathcal{N}}m_{\Lambda_b}^2 \int_{m_{\Lambda_b}}^{s_0} ds e^{-s/\tau} (s-m_{\Lambda_b}(x_2+x_3)/2)^3,
\end{eqnarray}
where the relevant parameters,  Gegenbauer moment $a_2$, dimensionful parameters $\epsilon_i(i=0,1,3)$, Borel mass $\tau$, continuum threshold $s_0$ and normalization constant $\mathcal{N}$ are used as
\begin{eqnarray}\label{eq:parameters1}
&&a_2=0.333^{+0.250}_{-0.333},\quad \epsilon_0=200^{+130}_{-60}\ \text{MeV},\quad \epsilon_1=650^{+650}_{-300}\ \text{MeV}, \quad \epsilon_3=230^{+60}_{-60}\ \text{MeV},\notag\\
&&\tau\in (0.4, 0.8) \text{GeV},\quad  s_0=1.2 \text{GeV},\quad \mathcal{N}=\int_{0}^{s_0} ds s^5 e^{-s/\tau}.
\end{eqnarray}

In this work, the $\Lambda_b$ baryon decays may proceed to final states containing either light pseudoscalar mesons or charmed meson. For a light pseudoscalar meson moving along the light-con direction $n$, its light-cone wave function is defined as~\cite{Li:2005kt},
\begin{eqnarray}
\langle P(p_2)|\bar q_{2\beta}(z) q_{1\alpha}(0)|0\rangle
&=&-\frac{i}{\sqrt{2N_c}}\int_0^1 dx e^{ixp_2\cdot z}\Bigg\{ \gamma_5 {p_2}\!\!\!\!/ \phi^{A}(x)+ m_0^P \gamma_5 \phi^P(x)-m_0^P \sigma_{\mu\nu} \gamma_5 p_{2}^{\mu} z^{\nu} \frac{\phi^{\sigma}(x)}{6}\Bigg\}_{\alpha\beta},\\
&=&-\frac{i}{\sqrt{2N_c}}\int_0^1 dx e^{ixp_2\cdot z}\Bigg\{ \gamma_5 {p_2}\!\!\!\!/ \phi^{A}(x)+ m_0^P \gamma_5 \phi^P(x)+m_0^P\gamma_5(\mathbf{n}\!\!\!/ \bar{\mathbf{n}}\!\!\!/-1) \phi^T(x)\Bigg\}_{\alpha\beta},
\end{eqnarray}
while the wave function of charmed D meson is defined by the light cone matrix element~\cite{Kurimoto:2002sb}:
\begin{eqnarray}
\int_{0}^1 \frac{d^4 z}{(2\pi)^4} e^{i k_2\cdot z} \langle 0|\bar c_{\alpha}(0) q_{\beta}(z)|\bar D^0(P_D))\rangle
&=&-\frac{i}{\sqrt{2N_c}}\Bigg\{ ({P\!\!\!\!/}_{D}+m_{0}^D)\gamma_5 \phi_{D}(k_2)\Bigg\}_{\beta\alpha}.
\end{eqnarray}
For the numerical calculation, we adopt the parametrization of distribution amplitude~\cite{Keum:2003js},
\begin{eqnarray}
&&\phi_{\pi}^A(x)=\frac{3f_{\pi}}{\sqrt6}x(1-x)(1+0.44 C_2^{3/2}(t)),\\
&&\phi_{\pi}^P(x)=\frac{f_{\pi}}{2\sqrt6}(1+0.43 C_2^{1/2}(t)),\\
&&\phi_{\pi}^T(x)=-\frac{f_{\pi}}{2\sqrt6}(C_1^{1/2}(t)+0.55 C_3^{1/2}(t)),\\
&&\phi_{K}^A(x)=\frac{3f_K}{\sqrt6}x(1-x)(1+0.17 C_1^{3/2}(t)+0.115 C_2^{3/2}(t)),\\
&&\phi_{K}^P(x)=\frac{f_K}{2\sqrt6}(1+0.24 C_2^{1/2}(t)),\\
&&\phi_K^T(x)=-\frac{f_K}{2\sqrt6}(C_1^{1/2}(t)+0.35 C_3^{1/2}(t)),\\
&&\phi_{D}(x)=\frac{f_{D}}{2\sqrt{2N_c}}6x(1-x)[1+C_D (1-2x)], 
\end{eqnarray}
here $t=2x-1$, and the Gegenbauer polynomials are shown as,
\begin{eqnarray}
&&C_1^{1/2}(t)=t,\ C_1^{3/2}(t)=3t,\ C_2^{1/2}(t)=\frac{1}{2}(3t^2-1),\ C_2^{3/2}(t)=\frac{3}{2}(5t^2-1),\notag\\
&&C_3^{1/2}(t)=\frac{1}{2}t(5t^2-3),\ C_4^{1/2}(t)=\frac{1}{8}(35t^4-30t^2+3),\ C_4^{3/2}(t)=\frac{15}{8}(21t^4-14t^2+1).
\end{eqnarray}
with the free shape parameters $C_D$, $\omega_D$~\cite{Kim:2013ria}, chiral enhancement parameter $m_0^P=m_P^2/(m_{q_1}+m_{q_2})$,  decay constants $f_{\pi}$, $f_K$, $f_D$~\cite{Aoki:2019cca} in our numerical calculations are adopted as,
\begin{eqnarray}\label{eq:parameters2}
&&C_D=0.5\pm0.1,\quad \omega_D=0.1,\quad m_0^{\pi}=1.4\pm 0.1\ \text{GeV},\quad m_0^K=1.6\pm 0.1\ \text{GeV},\quad  m_0^D=1.87\pm0.1\ \text{GeV},\notag\\
&&f_{\pi}=130\ \text{MeV},\quad f_K=160\ \text{MeV},\quad f_D=0.209\pm 0.002\ \text{GeV},
\end{eqnarray}

\section{The B-mesogenesis scenario}
\label{sec:bmesogenesis}
In B-Mesogenesis scenarios, the $b$-quark can decay into two light quarks and a dark baryon $\psi$. While the total baryon number is conserved in this process, the invisibility of dark baryon $\psi$ creates an apparent baryon number violation in the visible sector. This signature of baryon non-conservation emerges from the following two types effective Hamiltonian\cite{Alonso-Alvarez:2021qfd}:
\begin{eqnarray}\label{eq:hamiltonian}
&&\mathcal{H}_{eff}^{I}=-G^{I}_{uq} \bar{\mathcal{O}}^{I}_{uq} \psi^c+h.c. \ G^{I}_{uq}=\frac{y_{ub}y_{\phi q}}{M_Y^2},
\ \mathcal{O}^I_{uq}=-i \epsilon_{ijk}(u^{iT}C P_R b^j)P_R q^k,\notag\\
&&\mathcal{H}_{eff}^{II}=-G^{II}_{uq} \bar{\mathcal{O}}^{II}_{uq} \psi^c+h.c. \ G^{II}_{uq}=\frac{y_{\phi b}y_{u q}}{M_Y^2},
\ \mathcal{O}^{II}_{uq}=-i \epsilon_{ijk}(u^{iT}C P_R q^j)P_R b^k,
\end{eqnarray}
with $P_R=(1+\gamma_5)/2$ and $C$ is the charge conjugate matrix, $y_{ij}$ being the product of the two relevant dimensionless couplings, $M_Y$ is the mass of dark baryon. Using the effective Hamiltonians, the decay amplitude for $\Lambda_b\to M \psi$ can be expressed as,
\begin{eqnarray}
iM=-G_{uq} \bar u_{\psi}^c(q,s_{\psi}) \langle M(p_2)|\mathcal{O}_{uq}(0)|\Lambda_b(p_1,s_{\Lambda_b})\rangle,
\end{eqnarray}
where $u_{\psi}$ is the spinor of dark baryon, and momentum $q=p_1-p_2$. The transition matrix element on the right hand side above can be parameterized by two form factors $F_1(q^2)$ and $F_2(q^2)$,
\begin{eqnarray}\label{eq:1}
\langle M(p_2)|\mathcal{O}_{uq}(0)|\Lambda_b(p_1,s_{\Lambda_b})\rangle=P_R\Big[F_1(q^2)+\frac{q\!\!\!/}{m_{\Lambda_b}}F_2(q^2) \Big] u_{\Lambda_b}(p_1,s_{\Lambda_b}).
\end{eqnarray}
Using the preceding definitions, the decay width for the characteristic baryon-number-violating decays $\Lambda_b\to \pi \phi$, $\Lambda_b\to K\phi$, and $\Lambda_b\to D \phi$ are derived as follows,
\begin{eqnarray}
\Gamma(\Lambda_b\to M\psi)=\frac{G_{uq}^2 |\overrightarrow{q}|}{8 m_{\Lambda_b}^2(2\pi)^5}\Big[ (m_{\Lambda_b}^2+m_{\psi}^2-m_M^2)\Big(F_1^2(m_{\psi}^2)+\frac{m_{\psi}^2}{m_{\Lambda_b}^2}F_2^2(m_{\psi}^2)\Big) +4 m_{\psi}^2 F_1(m_{\psi}^2) F_2(m_{\psi}^2))\Big].
\end{eqnarray}
The kinematic variables $|\overrightarrow{q}|$  denote the magnitudes of the dark baryon $\psi$ momenta in the center-of-mass frame,
\begin{eqnarray}
|\overrightarrow{q}|=\frac{1}{2m_{\Lambda_b}}\sqrt{[m_{\Lambda_b}^2-(m_{\psi}+m_M)^2][m_{\Lambda_b}^2-(m_{\psi}-m_M)^2]}.
\end{eqnarray}
The coupling constants $G_{uq}$ can be constrained through $B$-meson semi-inclusive decays. Available theoretical determinations using heavy quark operator product expansion(OPE) yield upper limits consistent with current experimental results\cite{Shi:2024uqs},
\begin{eqnarray}
&\text{Type\ I}:\quad&G_{ud}^2<(1.8\pm0.35)\times 10^{-14}\ \text{GeV}^{-4},\\
&&G_{us}^2<(3.75\pm0.74)\times 10^{-14}\ \text{GeV}^{-4},\\
&&G_{cd}^2<(1.06\pm0.21)\times 10^{-12}\ \text{GeV}^{-4},\\
&&G_{cs}^2<(1.63\pm0.33)\times 10^{-12}\ \text{GeV}^{-4},\\
&\text{Type\ II}:\quad &G_{us}^2<(1.07\pm0.21)\times 10^{-11}\ \text{GeV}^{-4},\\
&&G_{cs}^2<(3.62\pm0.72)\times 10^{-10}\ \text{GeV}^{-4},
\end{eqnarray}
In the numerical analysis, the dark baryon mass is restricted to the kinematic range $0<m_{\psi}<m_{\Lambda_b}-m_M$, while the couplings $G_{ij}$ are fixed at their center values of upper limit.
For the type-II scheme, an additional four-fermion coupling is required in this work, parameterized as $|G_{ud}|^2=10^{-13}\ \text{GeV}^{-4}$~\cite{Khodjamirian:2022vta}.
\section{Perturbative Calculations}
\label{sec:perturbative calculations}
The process $\Lambda_b \to M \psi$ can be handled by factorization theorems. Usually, the factorization formula for the nonleptonic hadronic decays can be expressed as
\begin{eqnarray}
\mathcal{M}\sim \int_0^1 dx_2dx_3dx_4 \int d^2\mathbf{b}_2 d^2\mathbf{b}_3 d^2 \mathbf{b}_4\; G_{ij}(t) \phi_{\Lambda_b}(x_2,x_3,\mathbf{b}_2,\mathbf{b}_3,t) H(x_2,x_3,x_4,\mathbf{b}_2,\mathbf{b}_3,\mathbf{b}_4,t)\phi_{M}(x_4,\mathbf{b}_4,t),
\end{eqnarray}
where the coefficients $C_{ij}(t)$, organizing the QCD corrections for the effective operator in Eq.\ref{eq:hamiltonian}. The hard kernel $H(x_i,\mathbf{b}_i,t)$, representing $b$-quark decay sub-amplitude, and the nonperturbative meson wave function $\phi_{i}(x_i,\mathbf{b}_i,t)$, describes the evolution from scale $t$ to the lower hadronic scale $\Lambda_{QCD}$. For a review of this approach, see Ref.~\cite{Li:2003yj}.

We will work in the light-cone coordinates. The initial $\Lambda_b$ baryon is assumed to be at rest, with the final meson recoiling along the light-cone direction $\mathbf{n}$. Then the momentums of hadrons and their valence quarks are parametrized as follows,
\begin{eqnarray}
&&p_1(\Lambda_b)=\frac{m_{\Lambda_b}}{\sqrt2}(1,1,\mathbf{0_\perp}),\ p_2(M)=\frac{m_{\Lambda_b}}{\sqrt2}(\eta,0,\mathbf{0_\perp}),\\
&&k_1=\frac{m_{\Lambda_b}}{\sqrt2}(1,x_1,\mathbf{k_{1\perp}}),\ k_2=\frac{m_{\Lambda_b}}{\sqrt2}(0,x_2,\mathbf{k_{2\perp}}),\ k_3=\frac{m_{\Lambda_b}}{\sqrt2}(0,x_3,\mathbf{k_{3\perp}}),\\
&&k_4=\frac{m_{\Lambda_b}}{\sqrt2}(x_4 \eta,0,\mathbf{k_{4\perp}}),\; k_5=\frac{m_{\Lambda_b}}{\sqrt2}(\bar x_4 \eta, 0, -\mathbf{k_{4\perp}} ).
\end{eqnarray}
where $p_1$ and $p_2$ denote the momentum of the $\Lambda_b$ baryon and final meson, $k_1$ is the $b$-quark momentum, $k_2$ and $k_3$ are the two light quark momentum in $\Lambda_b$, $k_4$ and $k_5$ are quark momentums in final meson. $x_i$ are their longitudinal momentum fractions, and $\mathbf{k_{i\perp}}$ are the corresponding transverse momenta, satisfying $\Sigma_{i=1,2,3} \mathbf{k_{i\perp}}=0$.
Accordingly, the transfer momentum and light-cone components can be achieved as $q^2=(p_{1}-p_{2})^2=(1-\eta)m_{\Lambda_b}^2$. In the heavy quark limit, $m_{\Lambda_b}=m_{b} +\bar \Lambda$($\bar \Lambda$ is the order of QCD scale). Since $m_{\Lambda_b}\gg \bar \Lambda$, we expand the amplitudes in terms of $\frac{\bar \Lambda}{m_{\Lambda_b}}$. At the leading order of expansion, $\eta\sim1, q^2\sim0$.

\begin{figure}
  \centering
  \includegraphics[width=0.99\columnwidth]{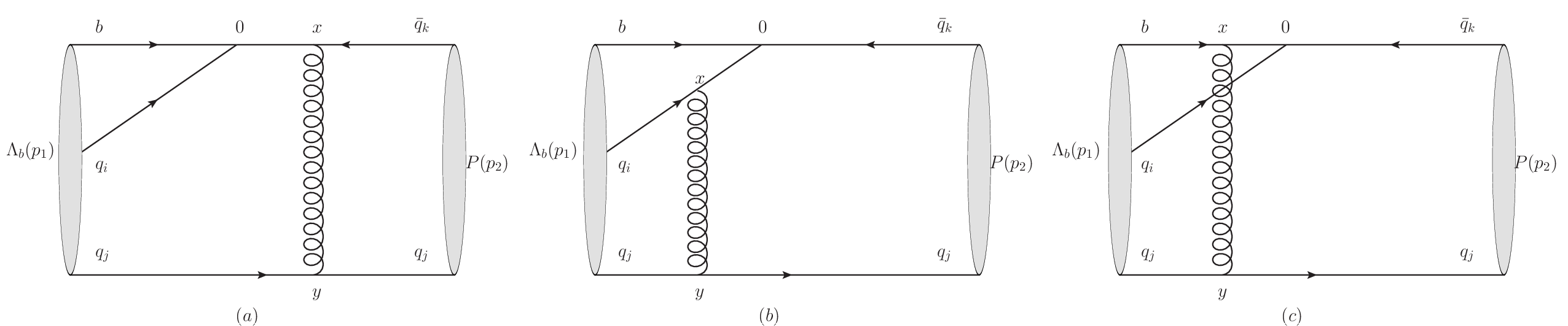}\\
  \caption{Typical leading order Feynman diagrams for the production of dark baryon with the bottom-baryon decays $\Lambda_b\to P\psi$. $q_i, q_j$ are the quarks in initial $\Lambda_b$ baryon. The final meson $P$ can be chosen as $\pi, K, D$.}\label{fig:fig1}
\end{figure}

In $k_T$-factorization, the three leading order feynman diagrams for $\Lambda_b\to M (\psi)$ decays are shown in Fig.~\ref{fig:fig1}. The corresponding decay amplitudes take two distinct contraction forms $\mathcal{M}_x(q_i,q_j,\bar q_k)=\mathcal{M}_x^{ud}(u,d,\bar d/\bar s)+\mathcal{M}_x^{du}(d,u,\bar u/\bar c)$. For the Type-I effective Hamiltonian, these amplitudes are explicitly given by,
\begin{eqnarray}
\mathcal{M}_{a}&=&\frac{2\pi C_F m_{\Lambda_b}}{9\sqrt3} E(t_a)\int^1_0dx_3dx_4\int^{1/\Lambda}_0\mathbf{b_3}d\mathbf{b_3}\mathbf{b_4}d\mathbf{b_4} h_1(x_3,x_4,\mathbf{b_3},\mathbf{b_4})P_R\Bigg\{ \bigg[ -\sqrt2 m_0m_{\Lambda_b}\eta f_{\Lambda_b}^{(2)}\phi^P(x_4,\mathbf{b_4})\phi_4(x_3,\mathbf{b_3})\notag\\
&&+n\!\!\!/\Big(m_0m_{\Lambda_b}f_{\Lambda_b}^{(2)}\phi^P(x_4,\mathbf{b_4})(x_3\phi_2(x_3,\mathbf{b_3})+\eta \phi_4(x_3,\mathbf{b_3}))
-\frac{\sqrt2}{4}m_{\Lambda_b}^2 x_3 \eta f_{\Lambda_b}^{(1)}\phi^A(x_4,\mathbf{b_4})\phi_3^{+-}(x_3,\mathbf{b_3}) \Big)\bigg]^{ud}\notag\\
&&+\bigg[ \sqrt2m_0 m_{\Lambda_b} x_3 f_{\Lambda_b}^{(2)} \phi^P(x_4,\mathbf{b_4})\phi_2(x_3,\mathbf{b_3}) +\frac{1}{2}m_{\Lambda_b}^2 \eta x_3 f_{\Lambda_b}^{(1)} \phi^A(x_4,\mathbf{b_4})\phi_3^{-+}(x_3,\mathbf{b_3})+ n\!\!\!/\Big( -\frac{\sqrt2}{4}m_{\Lambda_b}^2 x_3 \eta f_{\Lambda_b}^{(1)}\notag\\
&& \phi^A(x_4,\mathbf{b_4}) \phi_3^{-+}(x_3,\mathbf{b_3})-m_0 m_{\Lambda_b} f_{\Lambda_b}^{(2)} \Phi^P(x_4,\mathbf{b_4})(x_3 \phi_2(x_3,\mathbf{b_3}) +\eta \phi_4(x_3,\mathbf{b_3}) )\Big)   \bigg]^{du} \Bigg\}u_{\Lambda_b},\\
\mathcal{M}_{b}&=&\frac{\pi C_F m_{\Lambda_b}^2}{9\sqrt6} E(t_b)\int^1_0dx_2dx_3dx_4\int^{1/\Lambda}_0\mathbf{b_2}d\mathbf{b_2}\mathbf{b_3}d\mathbf{b_3} \mathbf{b_4}d\mathbf{b_4} h_2(x_2,x_3,x_4,\mathbf{b_2},\mathbf{b_3},\mathbf{b_4})P_R\Bigg\{ \bigg[ \frac{\sqrt2}{2}m_0m_{\Lambda_b}(x_2+x_3)f_{\Lambda_b}^{(2)}\notag\\
&&\phi_2(x_2,x_3,\mathbf{b_2},\mathbf{b_3})(\phi^T(x_4,\mathbf{b_4})-\phi^P(x_4,\mathbf{b_4}))-\frac{1}{2}m_{\Lambda_b}^2 \eta (x_2+x_3)f_{\Lambda_b}^{(1)}
\phi^A(x_4,\mathbf{b_4})(\phi_3^{+-}(x_2,x_3,\mathbf{b_2},\mathbf{b_3})\notag\\
&&+\phi_3^{-+}(x_2,x_3,\mathbf{b_2},\mathbf{b_3}))+n\!\!\!/\Big(\frac{1}{2}m_0m_{\Lambda_b}\eta x_4 f_{\Lambda_b}^{(2)} \phi_4(x_2,x_3,\mathbf{b_2},\mathbf{b_3})(\phi^P(x_4,\mathbf{b_4})+\phi^T(x_4,\mathbf{b_4}))\notag\\
&&+\frac{1}{2}m_0 m_{\Lambda_b} (x_2+x_3) f_{\Lambda_b}^{(2)} \phi_2(x_2,x_3,\mathbf{b_2},\mathbf{b_3})(\phi^P(x_4,\mathbf{b_4})-\phi^T(x_4,\mathbf{b_4}))+\frac{\sqrt2}{4}m_{\Lambda_b}^2\eta (x_2+x_3)f_{\Lambda_b}^{(1)}\phi^A(x_4,\mathbf{b_4})\notag\\
&&(\phi_3^{+-}(x_2,x_3,\mathbf{b_2},\mathbf{b_3})+\phi_3^{-+}(x_2,x_3,\mathbf{b_2},\mathbf{b_3})) \Big) \bigg]^{ud}+\bigg[\frac{\sqrt2}{2}m_0 m_{\Lambda_b}\eta x_4 f_{\Lambda_b}^{(2)}(\phi^P(x_4,\mathbf{b_4})+\phi^T(x_4,\mathbf{b_4}))\notag\\
&&\phi_4(x_2,x_3,\mathbf{b_2},\mathbf{b_3})+n\!\!\!/\Big(-\frac{1}{2} x_4 \eta m_0 m_{\Lambda_b} f_{\Lambda_b}^{(2)}\phi_4(x_2,x_3,\mathbf{b_2},\mathbf{b_3})(\phi^P(x_4,\mathbf{b_4})+\phi^T(x_4,\mathbf{b_4})) -\frac{1}{2}m_0 m_{\Lambda_b}(x_2+x_3) f_{\Lambda_b}^{(2)}\nonumber\\
&& \phi_2(x_2,x_3,\mathbf{b_2},\mathbf{b_3}) (\phi^P(x_4,\mathbf{b_4})-\phi^T(x_4,\mathbf{b_4}))-\frac{1}{2}(\phi_3^{+-}(x_2,x_3,\mathbf{b_2},\mathbf{b_3})+\phi_3^{-+}(x_2,x_3,\mathbf{b_2},\mathbf{b_3}))\notag\\
&&(x_2+ x_3) \eta m_{\Lambda_b}^2 f_{\Lambda_b}^{(1)}\phi^A(x_4,\mathbf{b_4}) \Big) \bigg]^{du} \Bigg\} u_{\Lambda_b},\\
\mathcal{M}_{c}&=&\frac{\pi C_F m_{\Lambda_b}^2}{9\sqrt6} E(t_c)\int^1_0dx_2dx_3dx_4\int^{1/\Lambda}_0\mathbf{b_2}d\mathbf{b_2}\mathbf{b_3}d\mathbf{b_3} \mathbf{b_4}d\mathbf{b_4} h_3(x_2,x_3,x_4,\mathbf{b_2},\mathbf{b_3},\mathbf{b_4})P_R\Bigg\{ \bigg[\frac{1}{2}\bar x_2 \eta m_{\Lambda_b}^2 f_{\Lambda_b}^{(1)}\notag\\
&& \phi^A(x_4,\mathbf{b_4})\phi_3^{+-}(x_2,x_3,\mathbf{b_2},\mathbf{b_3})+\frac{\sqrt2}{2} m_0 m_{\Lambda_b} f_{\Lambda_b}^{(2)} (\bar x_2 \phi_2(x_2,x_3,\mathbf{b_2},\mathbf{b_3})+(1-x_4 \eta) \phi_4(x_2,x_3,\mathbf{b_2},\mathbf{b_3}))\notag\\
&&(\phi^P(x_4,\mathbf{b_4})-\phi^T(x_4,\mathbf{b_4})) +n\!\!\!/\Big( m_0m_{\Lambda_b} f_{\Lambda_b}^{(2)} \phi^T(x_4,\mathbf{b_4})(\bar x_2 \phi_2(x_2,x_3,\mathbf{b_2},\mathbf{b_3})+(1-x_4 \eta) \phi_4(x_2,x_3,\mathbf{b_2},\mathbf{b_3}) )\notag\\
&&-\frac{\sqrt2}{4}\bar x_2 \eta m_{\Lambda_b}^2 f_{\Lambda_b}^{(1)}\phi_3^{+-}(x_2,x_3,\mathbf{b_2},\mathbf{b_3}) \phi^A(x_4,\mathbf{b_4})
\Big)\bigg]^{ud}+\bigg[ \frac{1}{2}m_{\Lambda_b}^2 \eta \bar x_3 f_{\Lambda_b}^{(1)}\phi^A(x_4,\mathbf{b_4}) \phi_3^{-+}(x_2,x_3,\mathbf{b_2},\mathbf{b_3})\notag\\
&&+\frac{\sqrt2}{2} m_0m_{\Lambda_b}\bar x_3 f_{\Lambda_b}^{(2)}\phi_2(x_2,x_3,\mathbf{b_2},\mathbf{b_3})(\phi^P(x_4,\mathbf{b_4})-\phi^T(x_4,\mathbf{b_4}))+n\!\!\!/\Big(\frac{1}{2}m_0m_{\Lambda_b} \bar x_3 f_{\Lambda_b}^{(2)} \phi_2(x_2,x_3,\mathbf{b_2},\mathbf{b_3})  \notag\\
&& (\phi^T(x_4,\mathbf{b_4})-\phi^P(x_4,\mathbf{b_4}))+\frac{1}{2}m_0m_{\Lambda_b} (1-\eta x_4)f_{\Lambda_b}^{(2)}\phi_4(x_2,x_3,\mathbf{b_2},\mathbf{b_3})(\phi^P(x_4,\mathbf{b_4})+\phi^T(x_4,\mathbf{b_4}))\notag\\
&&+\frac{\sqrt2}{4} m_{\Lambda_b}^2 \eta \bar x_3  f_{\Lambda_b}^{(1)}\phi^A(x_4,\mathbf{b_4})(\phi_3^{+-}(x_2,x_3,\mathbf{b_2},\mathbf{b_3})-\phi_3^{-+}(x_2,x_3,\mathbf{b_2},\mathbf{b_3})) \Big) \bigg]^{du} \Bigg\}u_{\Lambda_b},
\end{eqnarray}
where $C_F$ is the color factor, the indices $ud$ and $du$ denote the contraction sequences of the valence quarks $q_i q_j$ within the initial $\Lambda_b$ baryon in Fig.~\ref{fig:fig1}. For the $\pi^0(\psi)$ final state, the decay amplitudes combines contributions from $-\frac{1}{\sqrt2}\mathcal{M}_x^{ud}(u,d,\bar d)$ and $\frac{1}{\sqrt2}\mathcal{M}_x^{du}(d,u,\bar u)$. In contract, the $K^0(\psi)$ and $\bar D^0(\psi)$ final states receive single-contribution amplitudes $\mathcal{M}_x^{ud}(u,d,\bar s)$ or $\mathcal{M}_x^{du}(d,u,\bar c)$ respectively. The corresponding amplitudes derived from the Type-II effective Hamiltonian are:
\begin{eqnarray}
\mathcal{M}_{a}'&=&-\frac{2\pi C_F m_{\Lambda_b}}{9\sqrt3} E(t_a)\int^1_0dx_3dx_4\int^{1/\Lambda}_0\mathbf{b_3}d\mathbf{b_3}\mathbf{b_4}d\mathbf{b_4} h_1(x_3,x_4,\mathbf{b_3},\mathbf{b_4})P_R\Bigg\{ \bigg[\sqrt{2}m_0 m_{\Lambda_b} f_{\Lambda_b}^{(2)}\phi^P(x_4,\mathbf{b_4}) (x_3 \phi_2(x_3,\mathbf{b_3})\notag\\
&&- \eta \phi_4(x_3,\mathbf{b_3}))+\frac{1}{2} m_{\Lambda_b}^2 x_3 \eta f_{\Lambda_b}^{(1)}\phi^A(x_4,\mathbf{b_4}) \phi_3^{+-}(x_3,\mathbf{b_3})\bigg]^{ud}+\bigg[\sqrt{2}m_0 m_{\Lambda_b} f_{\Lambda_b}^{(2)}\phi^P(x_4,\mathbf{b_4}) (x_3 \phi_2(x_3,\mathbf{b_3})\notag\\
&&- \eta \phi_4(x_3,\mathbf{b_3}))+\frac{1}{2} m_{\Lambda_b}^2 x_3 \eta f_{\Lambda_b}^{(1)}\phi^A(x_4,\mathbf{b_4}) \phi_3^{-+}(x_3,\mathbf{b_3})\bigg]^{du} \Bigg\}u_{\Lambda_b},\\
\mathcal{M}_{b}'&=&-\frac{\pi C_F m_{\Lambda_b}^2}{9\sqrt6} E(t_b)\int^1_0dx_2dx_3dx_4\int^{1/\Lambda}_0\mathbf{b_2}d\mathbf{b_2}\mathbf{b_3}d\mathbf{b_3}\mathbf{b_4}d\mathbf{b_4} h_1(x_2,x_3,x_4,\mathbf{b_2},\mathbf{b_3},\mathbf{b_4})P_R\Bigg\{\bigg[\frac{\sqrt2}{2}(x_2+x_3) m_0 m_{\Lambda_b} f_{\Lambda_b}^{(2)}\notag\\
&&\phi_2(x_2,x_3,\mathbf{b_2},\mathbf{b_3})(\phi^T(x_4,\mathbf{b_4})-\phi^P(x_4,\mathbf{b_4}))-\frac{\sqrt2}{2}\eta x_4 m_0 m_{\Lambda_b} f_{\Lambda_b}^{(2)} \phi_4(x_2,x_3,\mathbf{b_2},\mathbf{b_3}) (\phi^P(x_4,\mathbf{b_4})+\phi^T(x_4,\mathbf{b_4}))\notag\\
&&-\frac{1}{2}(x_2+x_3)\eta m_{\Lambda_b}^2 f_{\Lambda_b}^{(1)}\phi^A(x_4,\mathbf{b_4})(\phi_3^{+-}(x_2,x_3,\mathbf{b_2},\mathbf{b_3})+\phi_3^{-+}(x_2,x_3,\mathbf{b_2},\mathbf{b_3})) \bigg]^{ud}+\bigg[-\frac{\sqrt2}{2}(x_2+x_3) m_0 m_{\Lambda_b} f_{\Lambda_b}^{(2)}\notag\\
&& \phi_2(x_2,x_3,\mathbf{b_2},\mathbf{b_3})(\phi^P(x_4,\mathbf{b_4})-\phi^T(x_4,\mathbf{b_4}))-\frac{\sqrt2}{2}\eta x_4 m_0 m_{\Lambda_b} f_{\Lambda_b}^{(2)} \phi_4(x_2,x_3,\mathbf{b_2},\mathbf{b_3})(\phi^P(x_4,\mathbf{b_4})+\phi^T(x_4,\mathbf{b_4}))\notag\\
&&-\frac{1}{2}(x_2+x_3)\eta m_{\Lambda_b}^2 f_{\Lambda_b}^{(1)}\phi^A(x_4,\mathbf{b_4})(\phi_3^{+-}(x_2,x_3,\mathbf{b_2},\mathbf{b_3})+\phi_3^{-+}(x_2,x_3,\mathbf{b_2},\mathbf{b_3})) \bigg]^{du} \Bigg\}u_{\Lambda_b},\\
\mathcal{M}_{c}'&=&-\frac{\pi C_F m_{\Lambda_b}^2}{9\sqrt6} E(t_c)\int^1_0dx_2dx_3dx_4\int^{1/\Lambda}_0\mathbf{b_2}d\mathbf{b_2}\mathbf{b_3}d\mathbf{b_3}\mathbf{b_4}d\mathbf{b_4} h_1(x_2,x_3,x_4,\mathbf{b_2},\mathbf{b_3},\mathbf{b_4})P_R\Bigg\{ \bigg[\frac{\sqrt2}{2}m_0 m_{\Lambda_b} (1-\eta x_4) f_{\Lambda_b}^{(2)}\notag\\
&&(\phi^P(x_4,\mathbf{b_4})-\phi^T(x_4,\mathbf{b_4}))\phi_4(x_2,x_3,\mathbf{b_2},\mathbf{b_3})+n\!\!\!/\Big(\frac{1}{2} m_0 m_{\Lambda_b}(1-\eta x_4) f_{\Lambda_b}^{(2)}(\phi^T(x_4,\mathbf{b_4}) -\phi^P(x_4,\mathbf{b_4})) \notag\\
&&\phi_4(x_2,x_3,\mathbf{b_2},\mathbf{b_3}) +\frac{1}{2}m_0 m_{\Lambda_b}\bar x_2 f_{\Lambda_b}^{(2)} \phi_2(x_2,x_3,\mathbf{b_2},\mathbf{b_3}) (\phi^P(x_4,\mathbf{b_4}) +\phi^T(x_4,\mathbf{b_4}))-\frac{\sqrt2}{4}m_{\Lambda_b}^2 \eta \bar x_2 f_{\Lambda_b}^{(1)} \phi^A(x_4,\mathbf{b_4})\notag\\
&& \phi_3^{-+}(x_2,x_3,\mathbf{b_2},\mathbf{b_3}) \Big)\bigg]^{ud} +\bigg[ \frac{\sqrt2}{2}m_0 m_{\Lambda_b} (1-\eta x_4) f_{\Lambda_b}^{(2)} (\phi^P(x_4,\mathbf{b_4})-\phi^T(x_4,\mathbf{b_4}))\phi_4(x_2,x_3,\mathbf{b_2},\mathbf{b_3})  \notag\\
&&+ n\!\!\!/\Big(\frac{1}{2} m_0 m_{\Lambda_b}(1-\eta x_4) f_{\Lambda_b}^{(2)}(\phi^T(x_4,\mathbf{b_4}) -\phi^P(x_4,\mathbf{b_4}))
\phi_4(x_2,x_3,\mathbf{b_2},\mathbf{b_3}) +\frac{1}{2}m_0 m_{\Lambda_b}\bar x_3 f_{\Lambda_b}^{(2)} \notag\\ &&\phi_2(x_2,x_3,\mathbf{b_2},\mathbf{b_3}) (\phi^P(x_4,\mathbf{b_4}) +\phi^T(x_4,\mathbf{b_4}))-\frac{\sqrt2}{4}m_{\Lambda_b}^2 \eta \bar x_3 \phi^A(x_4,\mathbf{b_4})\phi_3^{+-}(x_2,x_3,\mathbf{b_2},\mathbf{b_3})
\Big)\bigg]^{du}\Bigg\}u_{\Lambda_b}.
\end{eqnarray}
The forms for the offshellness of the intermediate gluon $\beta_{e_x}$ and quarks $\alpha_{e_x}$ in the $\Lambda_b\to M(\psi)$ process are denoted as,
\begin{eqnarray}
&&\alpha_{1}^{ud/du}=x_3 \eta m_{\Lambda_b}^2, \alpha_{2}^{ud}=x_4(x_2+x_3)\eta m_{\Lambda_b}^2, \alpha_{2}^{du}=x_4(x_2+x_3)\eta m_{\Lambda_b}^2,  \alpha_{3}^{ud}=\bar x_2 (x_4\eta-1) m_{\Lambda_b}^2, \alpha_{3}^{du}=\bar x_3 (x_4\eta-1) m_{\Lambda_b}^2,\nonumber\\
&&\beta_{1}^{ud/du}=x_3x_4 \eta m_{\Lambda_b}^2, \beta_{2}^{ud}=x_3 x_4 \eta m_{\Lambda_b}^2, \beta_{2}^{du}=x_2 x_4 \eta m_{\Lambda_b}^2,  \beta_{3}^{ud}=x_3\eta x_4 m_{\Lambda_b}^2, \beta_{3}^{du}=x_2\eta x_4 m_{\Lambda_b}^2,
\end{eqnarray}
The hard kernel functions $h_{i}(h_i^{\prime})$ are written as
\begin{eqnarray}
&&h_1(x_3,x_4,\mathbf{b_3},\mathbf{b_4})=S_t(x_3)\; K_0(\sqrt\alpha_1 |\mathbf{b_4-b_3}|)\; K_0(\sqrt{\beta_{1}}|\mathbf{b_4}|),\nonumber\\
&&h_2(x_2,x_3,x_4,\mathbf{b_2},\mathbf{b_3},\mathbf{b_4})=S_t(x_2x_4)\; K_0(\sqrt\alpha_2 |\mathbf{b_2}|)\; K_0(\sqrt{\beta_{2}}|\mathbf{b_2-b_3}|)|_{b_4=-b_2},\nonumber\\
&&h_3(x_2,x_3,x_4,\mathbf{b_2},\mathbf{b_3},\mathbf{b_4})=S_t(x_4)\; (\frac{i\pi}{2})^2H_0^{(1)}(\sqrt{|\alpha_{3}|}|\mathbf{b_2}|)\; H_0^{(1)}(\sqrt{|\beta_{3}|}|\mathbf{b_3}|)|_{b_4=b_2-b_3}.
\end{eqnarray}
where $K_0$ and $H_0=J_0+i Y_0$ are Bessel functions. $h_1'$ is identical to $h_1$, while $h_2'$ and $h_3'$ are obtained through the symmetry operation $\mathbf{b_2}\leftrightarrow \mathbf{b_3}$. The derivation of the relevant Fourier transformation proceeds as,
\begin{eqnarray}
&&\int d^2k\frac{e^{ik\cdot b}}{\alpha+k^2}=2\pi \big[K_0(\sqrt{\alpha}b)\theta(\alpha)+\frac{i \pi }{2} H_0(\sqrt{\alpha}b)\theta(-\alpha)\big].
\end{eqnarray}
The threshold resummation factor $S_t(x)$ follows standard parametrization:
\begin{eqnarray}
S_t(x)=\frac{2^{1+2c} \Gamma(3/2+c)}{\sqrt{\pi}\Gamma(1+c)}[x(1-x)]^c,
\end{eqnarray}
with the parameter $c=0.4$ in this paper.
The evolution factors $E_{x}(t)$s in the factorization formulas are given by
\begin{eqnarray}
E(t_a)&=&\alpha_s(t) {\rm exp}(-S_{\Lambda_b}(t)-S_{P}(t)),
\end{eqnarray}
where
\begin{eqnarray}
S_{\Lambda_b}(t)&=&s(x_2 m_{\Lambda_b}, \omega)+s(x_3 m_{\Lambda_b}, \omega)+\frac{8}{3}\int_{\omega}^t \frac{d\bar \mu}{\bar \mu}\gamma_{q}(\alpha_s(\bar \mu)),\nonumber\\
S_{M}(t)&=&s(x_4 \eta m_{\Lambda_b}, \mathbf{b_4})+s(\bar x_4 \eta m_{\Lambda_b},\mathbf{ b_4})+2\int_{1/\mathbf{b_3}}^t \frac{d\bar \mu}{\bar \mu}\gamma_{q}(\alpha_s(\bar \mu)),\\
S_{D}(t)&=&s(x_4\eta m_{\Lambda_b},\mathbf{b_4})+2\int_{1/\mathbf{b}_4}^t \frac{d\bar \mu}{\bar \mu} \gamma_{q}(\alpha_s(\bar \mu)).
\end{eqnarray}
The choices of $\omega$ is
\begin{eqnarray}
\omega=min(\frac{1}{b_1},\frac{1}{b_2},\frac{1}{b_3}),\ \ b_1=|\mathbf{b}_2-\mathbf{b}_3|, \ b_2=|\mathbf{b}_1-\mathbf{b}_3|,\ b_3=|\mathbf{b}_1-\mathbf{b}_2|,
\end{eqnarray}
with the quark anomalous dimension $\gamma_{q}=-\alpha_s/\pi$. The explicit expression of $s(Q,b)$ can be found, for example, in Appendix A of Ref~\cite{Ali:2007ff}. The hard scales are chosen as
\begin{eqnarray}
t_{a}&=&max(\sqrt{\alpha_{1}},\sqrt{\beta_{1}},\omega,1/\mathbf{b_4}),\;\\
t_b&=&max(\sqrt{\alpha_{2}},\sqrt{\beta_{2}},\omega,1/\mathbf{b_3},1/|\mathbf{b_2}-\mathbf{b_3}|),\\
t_c&=&max(\sqrt{\alpha_{3}},\sqrt{\beta_{3}},\omega,1/\mathbf{b_2},1/\mathbf{b_3}),
\end{eqnarray}
It is straightforward to extract the form factor $F_1$ and $F_2$ by comparing with Eq.(\ref{eq:1}), where $\mathcal{M}_i(n\!\!\!/)$ and $\mathcal{M}_i(I)$ correspond to the lorentz structures $n\!\!\!/$ and $I$ in the amplitudes.
\begin{eqnarray}
&&F_1^{\pi}(q^2=0)=\frac{1}{\sqrt2}\sum_i^{a,b,c}(\mathcal{M}^{du}_i(I)-\mathcal{M}^{ud}_i(I)
+\frac{1-\eta}{\eta}(\mathcal{M}^{du}_i(n\!\!\!/)-\mathcal{M}^{ud}_i(n\!\!\!/)),\\
&&F_2^{\pi}(q^2=0)=\sum_i^{a,b,c} \frac{1}{\eta}\mathcal{M}^{du}_i(n\!\!\!/)-\frac{1}{\eta} \mathcal{M}^{ud}_i(n\!\!\!/),\\
&&F_1^{K}(q^2=0)=\sum_i^{a,b,c}\mathcal{M}^{ud}_i(I)+\frac{\sqrt2(1-\eta)}{\eta}\mathcal{M}^{ud}_i(n\!\!\!/),\\
&&F_2^{K}(q^2=0)=\sum_i^{a,b,c}\frac{\sqrt2}{\eta} \mathcal{M}^{ud}_i(n\!\!\!/),\\
&&F_1^{D}(q^2=0)=\sum_i^{a,b,c}\mathcal{M}^{du}_i(I)+\frac{\sqrt2(1-\eta)}{\eta}\mathcal{M}^{du}_i(n\!\!\!/),\\
&&F_2^{D}(q^2=0)=\sum_i^{a,b,c}\frac{\sqrt2}{\eta} \mathcal{M}^{du}_i(n\!\!\!/).
\end{eqnarray}
\section{Numerical Results}
\label{sec:numerical results}
We take the QCD scale $\Lambda=0.25\pm0.05$, and the wave function parameters of in Eqs.~\ref{eq:parameters1} and ~\ref{eq:parameters2}. In addition, the transverse momentum dependent parton distributions of the final meson are also included. In the work, we use the Gaussian form to factorize the wave functions~\cite{Lu:2018obb},
\begin{eqnarray}
\phi(x,\mathbf{b})=\phi(x) exp(-\frac{\mathbf{b}^2}{4\beta^2}),
\end{eqnarray}
where $\phi(x)$ is the longitudinal momentum distribution amplitude, the exponential factor describes the transverse momentum distribution of final meson. The parameters $\beta$ characterize the shape of the transverse momentum distributions, which is expected at the order of QCD scale $\Lambda$. We choose $\beta^2=4$ GeV$^{-2}$.

\begin{figure}[htbp]
  \centering
  \begin{minipage}[t]{0.45\linewidth}
  \centering
  \includegraphics[width=0.9\columnwidth]{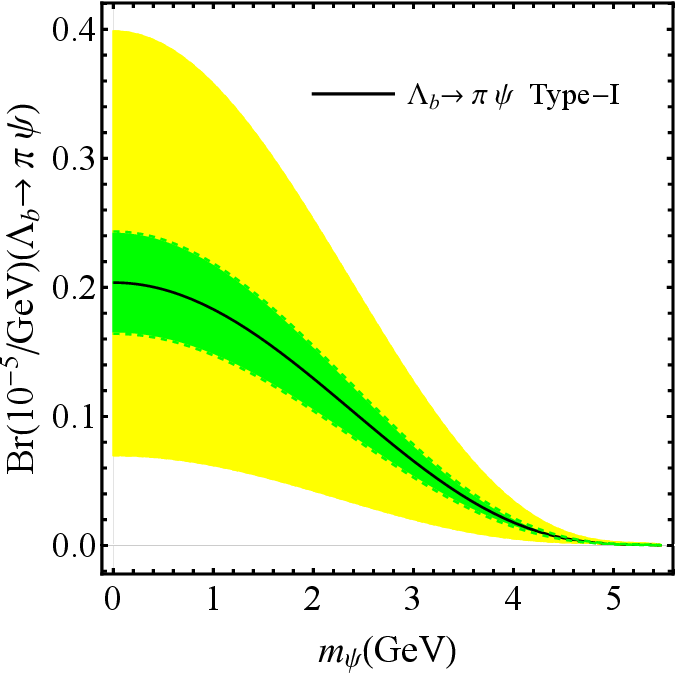}
  \end{minipage}
  \begin{minipage}[t]{0.45\linewidth}
  \centering
  \includegraphics[width=0.9\columnwidth]{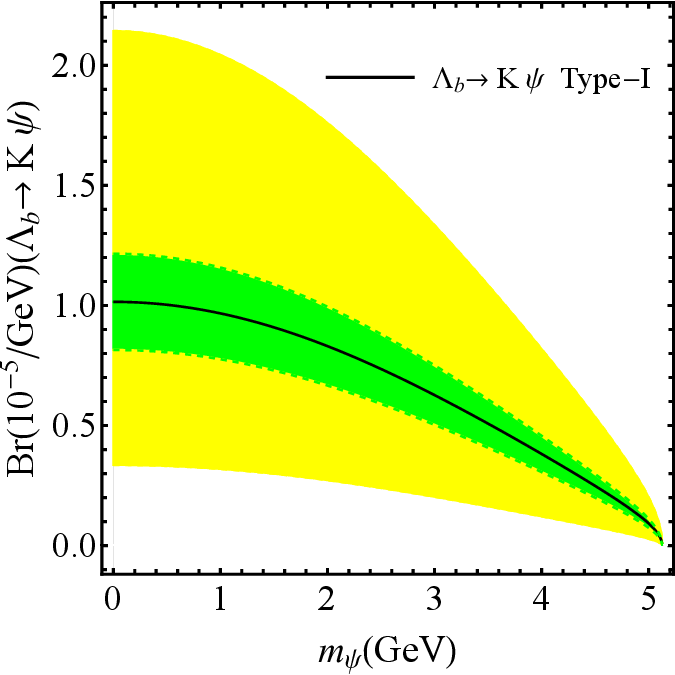}
  \end{minipage}
    \begin{minipage}[t]{0.45\linewidth}
  \centering
  \includegraphics[width=0.9\columnwidth]{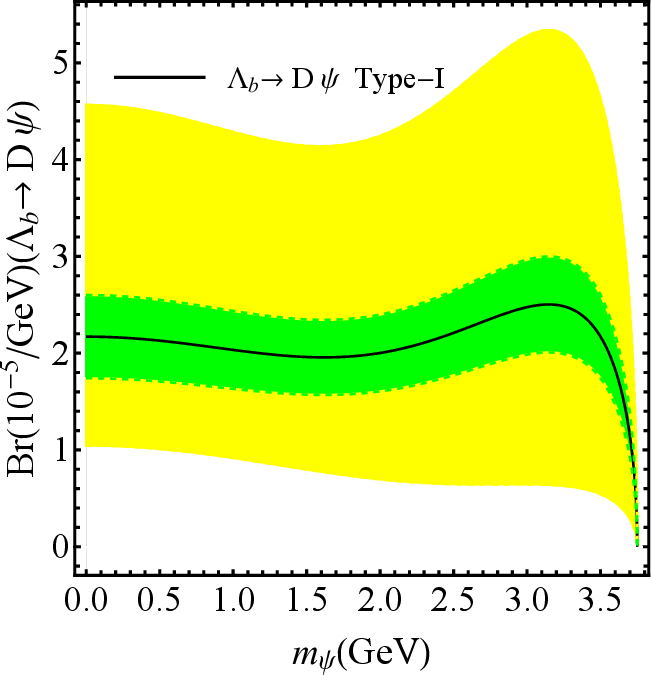}
  \end{minipage}
    \begin{minipage}[t]{0.45\linewidth}
  \centering
  \includegraphics[width=0.9\columnwidth]{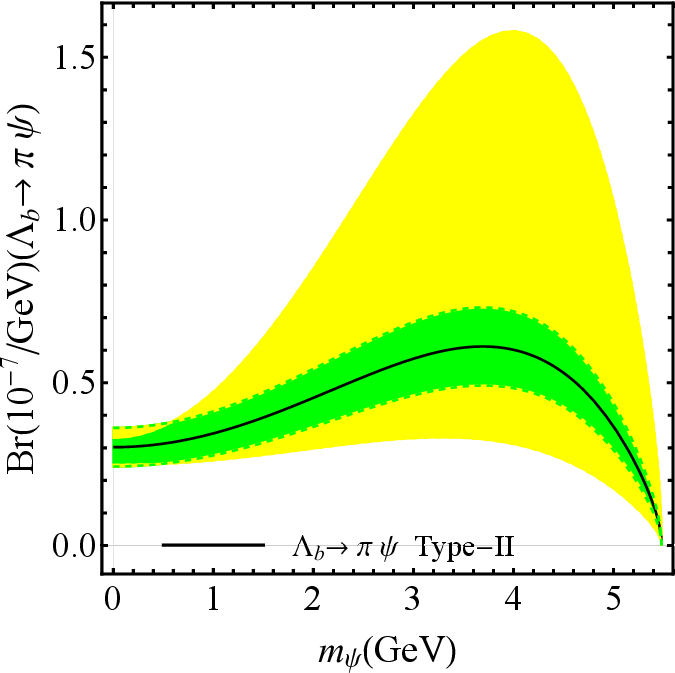}
  \end{minipage}
    \begin{minipage}[t]{0.45\linewidth}
  \centering
  \includegraphics[width=0.9\columnwidth]{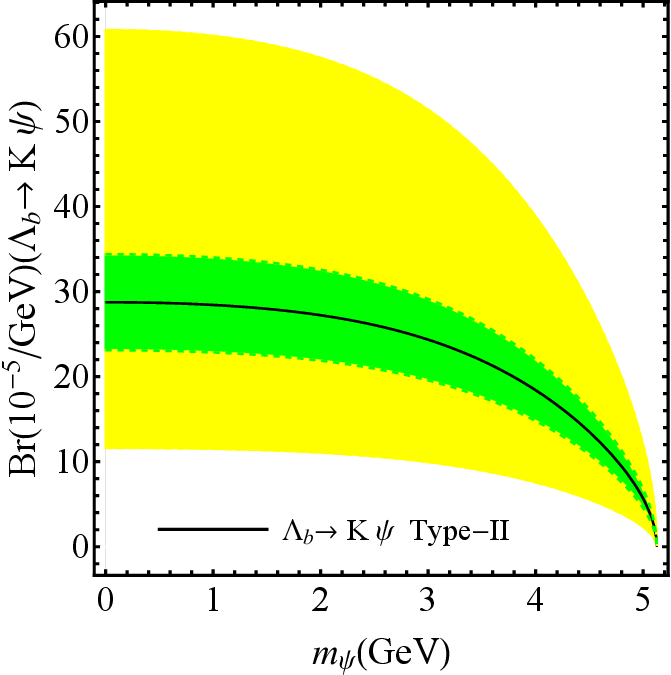}
  \end{minipage}
    \begin{minipage}[t]{0.45\linewidth}
  \centering
  \includegraphics[width=0.9\columnwidth]{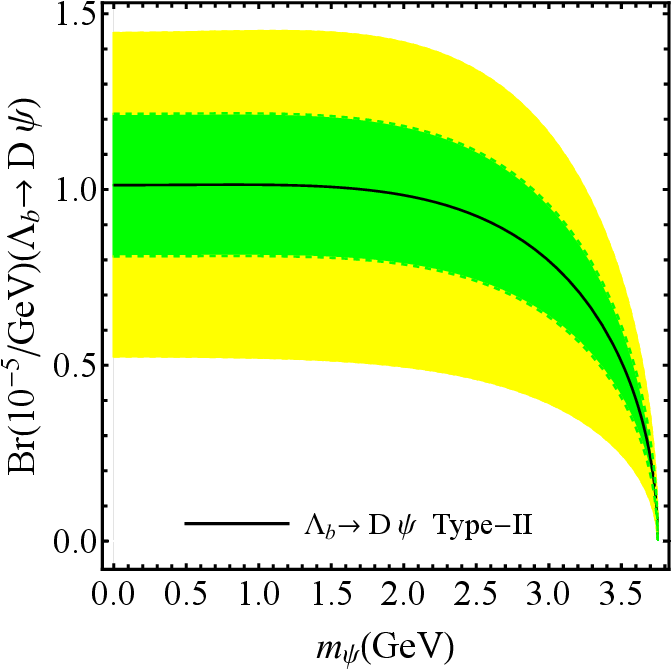}
  \end{minipage}
  \caption{The upper limits of branching fractions for $\Lambda_b\to M\psi$ as function of dark baryon mass $m_{\psi}$ in the Type-I and II models. The yellow and green bands denote the uncertainties from couplings $G_{ij}$ and form factor $F_{i}$.}\label{fig:fig2}
\end{figure}

Accordingly, the form factor $F_{1,2}(0)$ for bottom-baryon decays into meson can be obtained, the corresponding numerical results for the Type-I/II models are listed in Table.~\ref{tab:results}. The tabulated uncertainties primarily arise from the parameters of wave function and QCD scale $\Lambda$. These branching ratios are larger than those obtained via LCSR~\cite{Shi:2024uqs}, yet consistent with prediction in Ref~\cite{Alonso-Alvarez:2021qfd}. Furthermore, we present the full $m_{\psi}$-dependence of the predicted branching ratios in Fig.~\ref{fig:fig2}, showing upper limits for $\Lambda_b\to M\psi$ decays in Type I/II models. The yellow and green bands quantify uncertainties from couplings $G_{ij}$ and form factors $F_i$ respectively. Branching ratios reach $\mathcal{O}(10^{-5})$ in both models, with $\mathcal{B}(\Lambda_b\to K\psi)$ in Type II attaining $\mathcal{O}(10^{-4})$. These accessible magnitudes make our predictions testable at the LHCb and B-factories with improved precision. The pronounced sensitivity of branching ratios to the couplings $G_{ij}$, evident from the Fig.~\ref{fig:fig2}, motivates defining new ratios of different decay channels. For the Type I model, we define ratios $R_1^{I}$ and $R_2^{I}$,
\begin{eqnarray}
&&R_1^I=\frac{\mathcal{B}(\Lambda_b\to K^0 \psi)}{\mathcal{B}(\Lambda_b\to \pi^0 \psi)}\sim8.4,\quad R_2^I=\frac{\mathcal{B}(\Lambda_b\to \bar D^0 \psi)}{\mathcal{B}(\Lambda_b\to \pi^0 \psi)}\sim28.8,
\end{eqnarray}
with branching ratios at the benchmark dark baryon mass $m_{\psi}=\frac{1}{2}(m_{\Lambda_b}-m_{\pi})$. These ratios significantly deviates from the ratio of $G_{ij}$ factors:
\begin{eqnarray}
{R'}_{1}^I=\Big|\frac{y_{\psi s}}{y_{\psi d}}\Big|^2 \sim2.1,\quad {R'}_{2}^I=\Big|\frac{y_{cb}}{y_{ub}}\Big|^2 \sim58.9,\quad
\end{eqnarray}
which indicates large contribution at the amplitude level. For completeness, we obtain the upper limits of the $\Lambda_b\to M\psi$ branching fractions as functions of $m_{\psi}$ for Type I and II models, shown in Fig.~\ref{fig:fig2}. The yellow and green bands represent uncertainties from the couplings $G_{ij}$ and form factors $F_i$. Given the strong dependence of these parameters $G_{ij}$ and $F_{i}$, future measurements will provide essential constraints for exploring $\psi$-related physics.
\begin{table}
\caption{The form factors for the transition $\Lambda_b\to \pi(\psi),\ K(\psi),\ D(\psi)$ in the Type-I and II models.}\label{tab:results}
\begin{tabular}{c c cccccc}
\hline\hline
\multirow{2}{*}{ }& \multicolumn{2}{c}{Type-I} &\multirow{2}{*}{ }& \multicolumn{2}{c}{Type-II} \\
\cline{2-3}\cline{5-6}
&$F_1(0)$&$F_2(0)'$ &&$F_1(0)$&$F_2(0)'$\\
\hline
$\Lambda_b^0\to \pi^0$ & $1.189^{+0.474}_{-0.492}$ &$-1.400^{+0.440}_{-0.591}$&$\Lambda_b^0\to \pi^0$ & $0.059^{+0.003}_{-0.006}$& $0.055^{+0.082}_{-0.031}$\\
$\Lambda_b^0\to K^0$ &$-1.852^{+0.786}_{-0.839}$& $0.768^{+0.325}_{-0.305}$&$\Lambda_b^0\to K^0$ &$1.846^{+0.837}_{-0.672}$& $-0.146^{+0.071}_{-0.068}$ \\
$\Lambda_b^0\to \bar D^0$ & $0.569^{+0.256}_{-0.174}$& $-2.227^{+1.017}_{-0.948}$&$\Lambda_b^0\to \bar D^0$ & $1.264^{+0.246}_{-0.354}$& $0.107^{+0.051}_{-0.079}$\\
\hline
\hline
\end{tabular}
\end{table}


\section{Conclusions}
\label{sec:conclusions}
In this work, we studied the production of dark baryon $\psi$ by the decays of  $\Lambda_b$ within the PQCD framework for the first time. In the Type-I/II models, form factors and branching ratios of $\Lambda_b\to \pi \psi$, $\Lambda_b\to D\psi$ and $\Lambda_b\to K\psi$ are calculated respectively. The numerical results show that those branching ratios can reach the order of $10^{-5}$. The accessible magnitudes make our predictions testable at the LHCb and B-factories experiments. It should be pointed out that the coupling constants of the two types effective operators $G_{uq}$ may alter the results by a factor of two.


\end{document}